\documentclass[10pt, twocolumn, pre, aps, superscriptaddress, showpacs]{revtex4-1}
\usepackage{amsmath, graphicx, subfigure, tikz}
\begin{document}

\title{Multifractal Spin-Glass Chaos Projection and Interrelation of\\
Multicultural Music and Brain Signals}
    \author{E. Can Artun}
    \affiliation{Faculty of Engineering and Natural Sciences, Kadir Has University, Cibali, Istanbul 34083, Turkey}
\author{Ibrahim Ke\c{c}o\u{g}lu}
    \affiliation{Department of Physics, Bo\u{g}azi\c{c}i University, Bebek, Istanbul 34342, Turkey}
\author{Alpar T\"urko\u{g}lu}
    \affiliation{Department of Physics, Bo\u{g}azi\c{c}i University, Bebek, Istanbul 34342, Turkey}
    \affiliation{Department of Electrical and Electronics Engineering, Bo\u{g}azi\c{c}i University, Bebek, Istanbul 34342, Turkey}
\author{A. Nihat Berker}
    \affiliation{Faculty of Engineering and Natural Sciences, Kadir Has University, Cibali, Istanbul 34083, Turkey}
    \affiliation{Department of Physics, Massachusetts Institute of Technology, Cambridge, Massachusetts 02139, USA}

\begin{abstract}
A complexity classification scheme is developed from the fractal spectra of spin-glass chaos and demonstrated with multigeographic multicultural music and brain electroencephalogram signals.  Systematic patterns are found to emerge.  Chaos under scale change is the essence of spin-glass ordering and can be obtained, continuously tailor-made, from the exact renormalization-group solution of Ising models on frustrated hierarchical lattices.  The music pieces are from Turkish music, namely Arabesque, Rap, Pop, Classical, and Western music, namely Blues, Jazz, Pop, Classical.  A surprising group defection occurs.
\end{abstract}
\maketitle

\begin{figure}[ht!]
\centering
\includegraphics[scale=0.56]{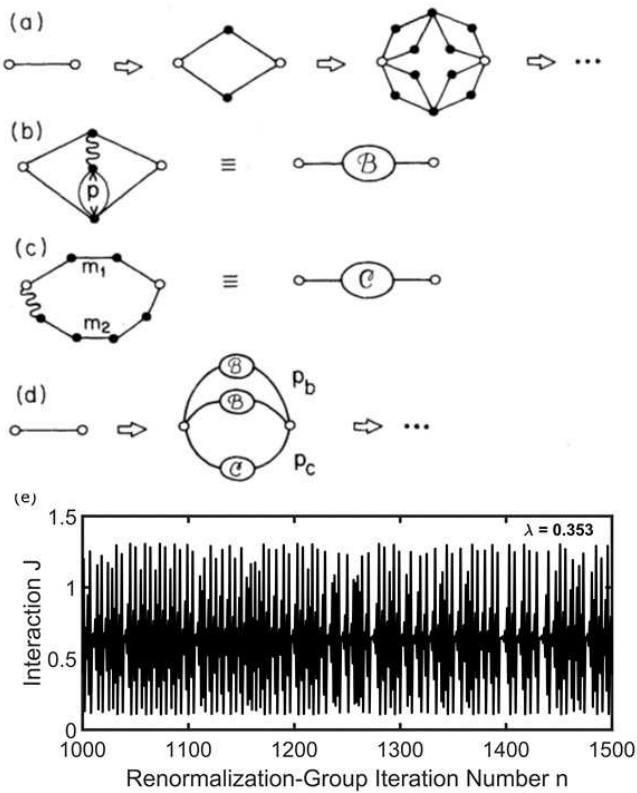}
\caption{Construction of the chaos hierarchical lattice, from Ref.\cite{McKayChaos}.  (a) Construction of a standard hierarchical lattice, as explained in Ref.\cite{BerkerOstlund}.  A graph is self-imbedded into each bond, \textit{ad infinitum}.  The exact renormalization-group solution proceeds in the reverse direction, by summing over the internal spins shown with the dark circles.  In (b) and (c) are shown two graphs generic to the microscopics of spin glasses.  The wiggly lines are infinite antiferromagnetic bonds, which have the sole effect of reversing the signs of the interactions adjoining on one side. (b) Frustrated units: Intermediate ferromagnetic (antiferromagnetic) ordering eliminates long-range antiferromagnetic (ferromagnetic) correlations.  (c) Depressed units: The correlation on the shortest path sustains, but is depressed by the competing longer path.  (d) Construction of the hierarchical lattice by combining the two generic units have yielded chaos in Ref.\cite{McKayChaos}. (e) A typical chaotic renormalization-group trajectory for this hierarchical lattice, \textit{e.g.,} occurring for index values $p=4,p_b=40,p_c=1,m_1=7,m_2=m_1+1$.  Each renormalization-group rescaling transformaton is a renormalization-group iteration, here consecutively denoted by $n$. For a given set of indices, renormalization-group trajectories, starting at any temperature within the spin-glass phase, fall to the same chaotic trajectory.  The Lyapunov exponent for the chaotic trajectory shown here is calculated (Eq.2) to be $\lambda = 0.353$.  The Lyapunov exponents $\lambda$ indicate the strength of chaos \cite{Gurleyen} and vary with the index values $p,p_b,p_c,m_1,m_2$.}
\end{figure}

\section{Introduction: Spin-Glass Chaos as a Complexity Classification Scheme}

One beauty of complex systems from different sources is similar properties hidden under the complicated behaviors, waiting to be unveiled. \cite{Eroglu1,Eroglu2} On the other hand, "Chaos under scale change" as the distinctive characteristic of a spin-glass phase \cite{McKayChaos,McKayChaos2,BerkerMcKay,Hartford,ZZhu,Katzgraber3,Fernandez,Fernandez2,Eldan,Wang2,Parisi3} and the multifractal spectrum quantification of exceedingly complicated data can be merged to create a classification scheme for complex systems.  In this scheme, the spin glasses can provide a standart metric for the wide range of complex systems.  The connections of the thus classified complex systems could then be  achieved, for example connecting different cultural trends, geographies, and time periods.

In the current work, we build such a complex-system classification scheme and illustrate its application with multicultural music and brain electroencephalogram (EEG) signals.
\begin{figure*}[ht!]
\centering
\includegraphics[scale=0.87]{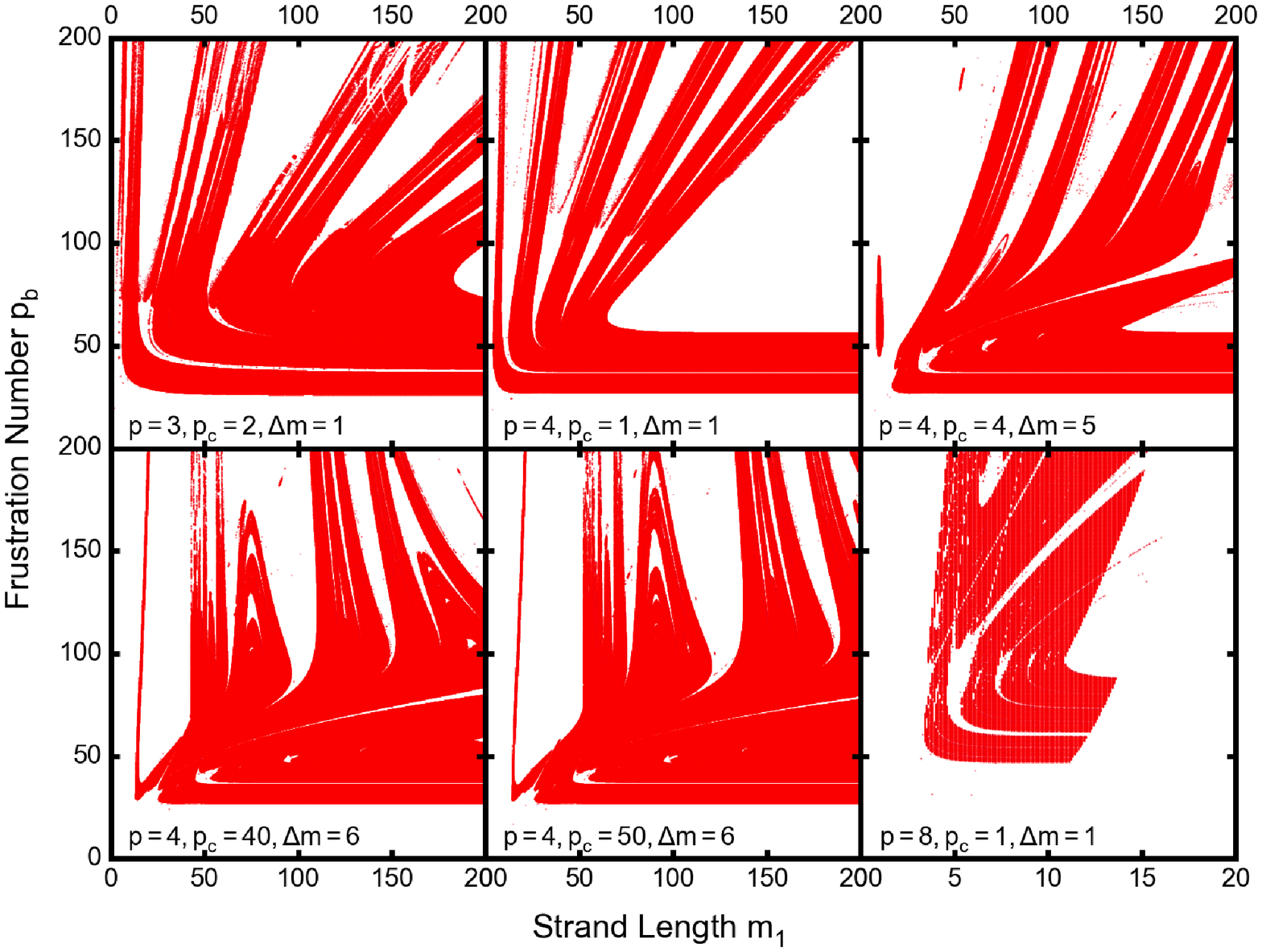}
\caption{Index regions of the frustrated hierarchical model that admit chaotic renormalization-group trajectories and, thus, a spin-glass phase.  Note islands that exclude chaos, such as the crescent and star region in the upper-right panel and all the lower panels.}
\end{figure*}
\begin{figure*}[ht!]
\centering
\includegraphics[scale=0.27]{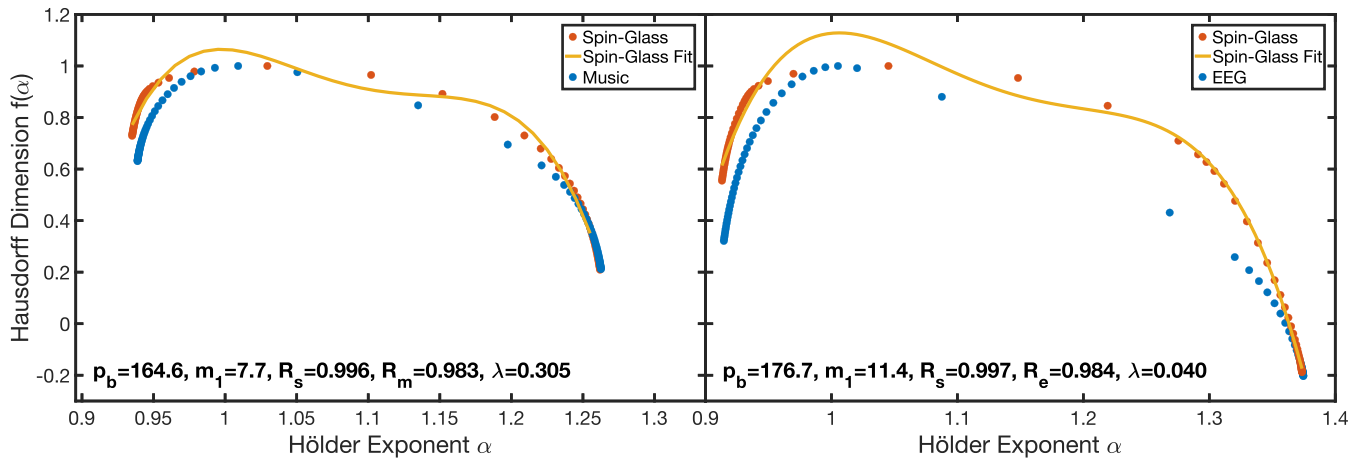}
\caption{Best fitting of SG multifractal spectrum points with the music spectrum (left panel) or brain EEG (right panel) points:  The SG multifractal spectrum points are fitted by a continuous quartic function.  The root-mean-square separation between the music or brain EEG spectrum points and the SG continuous function is minimized by varying the original spin-glass indices of frustration number $p_b$ and repressed strand length $m_1$.  At the end of the procedure, the correlation coefficient $R_s$ between the SG multifractal spectrum points and the SG continuous function, and the correlation coefficient $R_m$ or $R_e$ between the music spectrum or brain EEG points and SG continuous function $R_m$ together give the goodness of the fit between the SG multifractal spectrum points and the music or brain EEG spectrum points.  The calculated Lyapunov exponent $\lambda$ measures the strength of the fitting chaos. In this figure, the multifractal SG music fit is illustrated with Bul Beni, Ezhel (Table I) in the left panel and the multifractal SG brain EEG fit with Music Listening, cranular location $C_z$ (Fig.5) in the right panel.}
\end{figure*}
\begin{figure*}[ht!]
\centering
\includegraphics[scale=0.98]{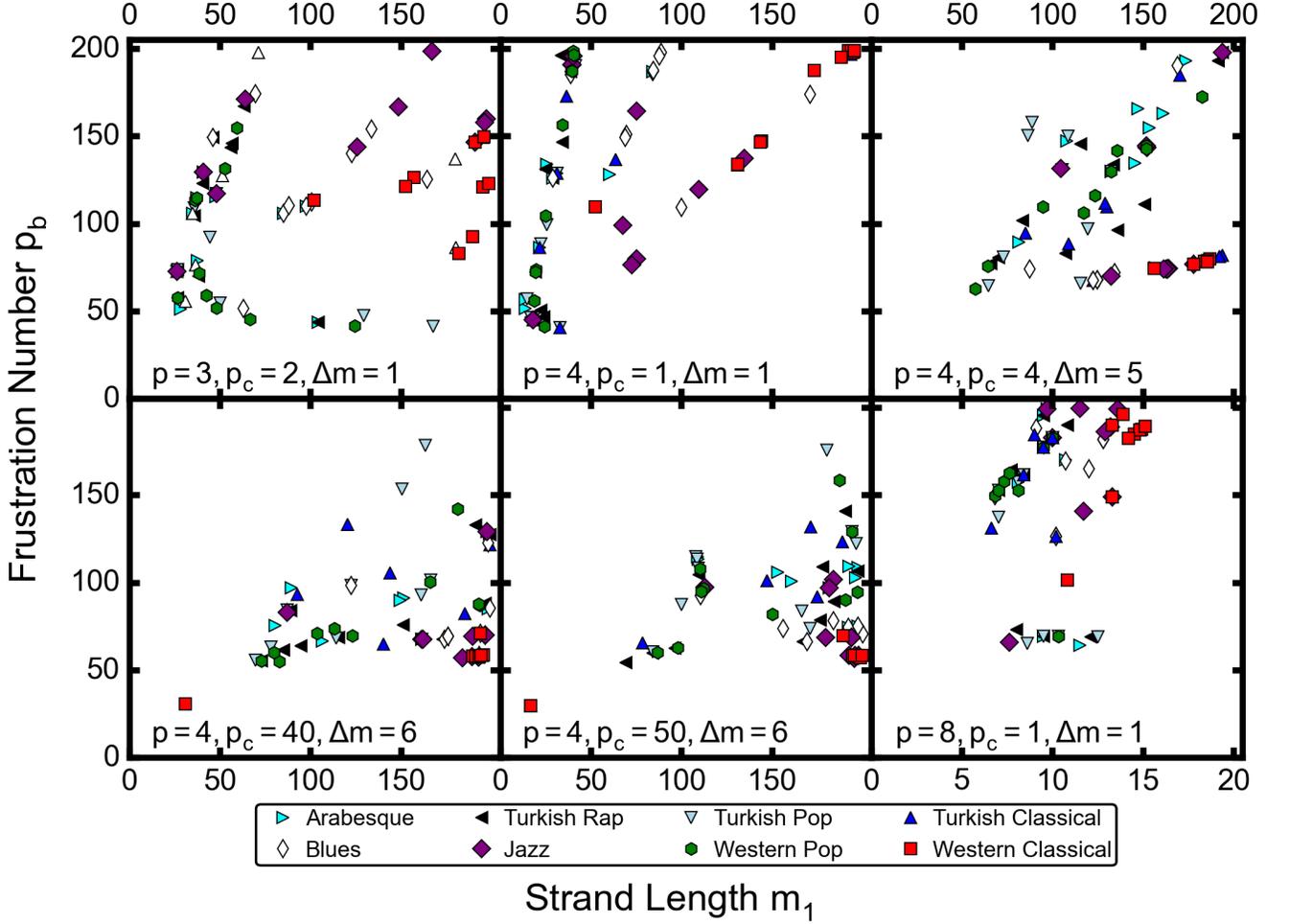} 
\caption{Scatter plots of the fitted frustrated hierarchical Ising models to multicultural music tracks. Each point represents the fitted model indices for a single music track. There are ten tracks, listed in Table I, for each of the eight genres.  In each panel, for the indicated values of $p,p_c,\Delta m$, the values of the frustration number $p_b$ and repressed strand length $m_1$ are fitted to the music tracks.  Systematic patterns emerge.}
\end{figure*}
\begin{figure*}[ht!]
\centering
\includegraphics[scale=0.52]{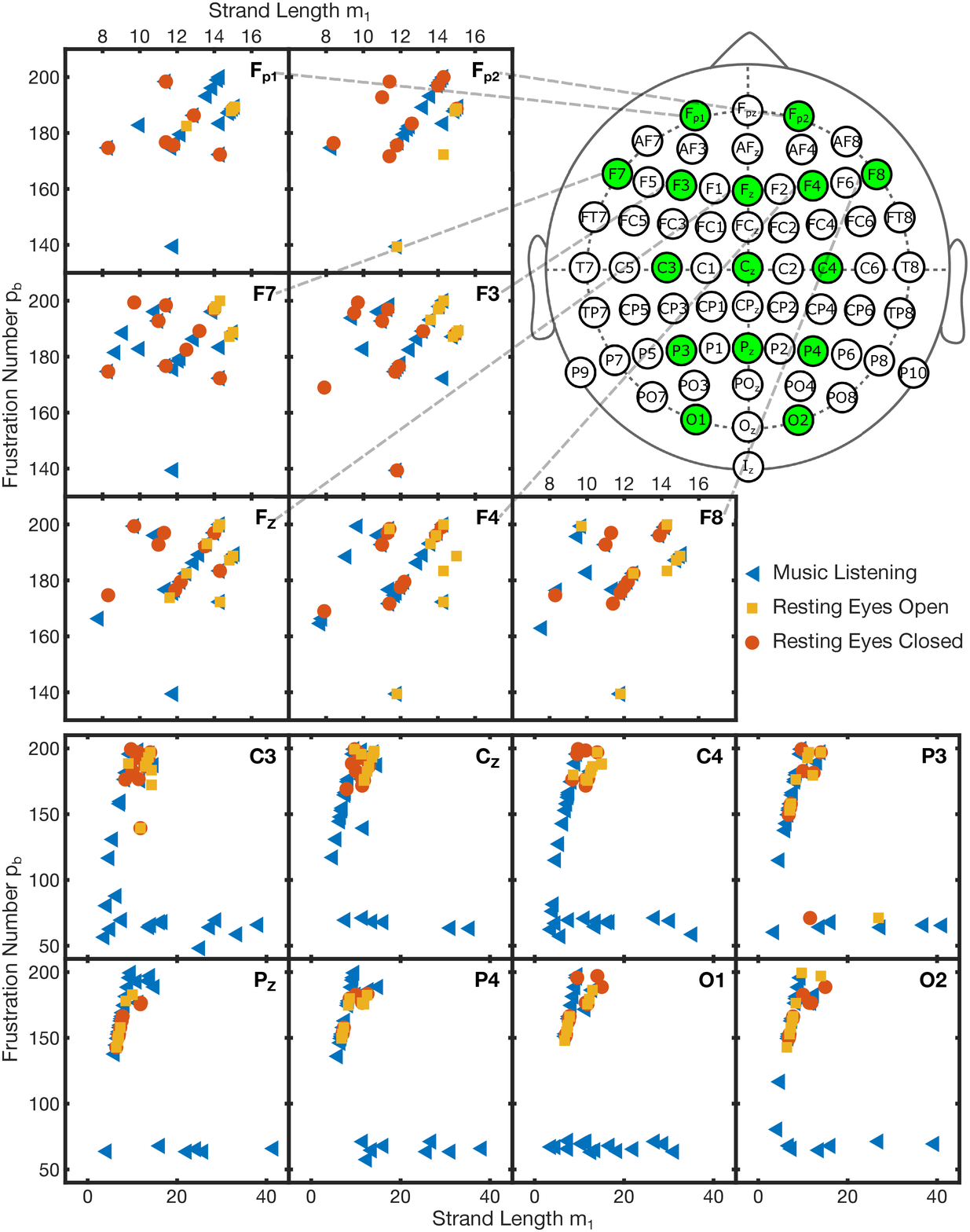}
\caption{SG multifractal parameters for EEG signals collected at different cranular probe locations, shown by the arrows, and in the different states of music listening, resting with eyes open or closed.  Again, an even clearer systematic pattern emerges.}
\end{figure*}
Frozen microscopic disorder introduced into an ordering system can immediately: Completely eliminate \cite{ImryMa} or reduce an ordered phase, change the critical exponents of a second-order phase transition \cite{Harris,AndelmanBerker}, convert a first-order phase transition into a second-order phase transition \cite{Aizenman,AizenmanE,HuiBerker,erratum}.  Competing microscopic interactions, with or without frozen microscopic disorder, furthermore can introduce a new phase, namely the spin-glass phase \cite{EdwardsAnderson}.  Competing microscopic interactions have two qualitatively different effects: (1) Two competing but non-cancelling (for example, due to unequal chain lengths) chains of interactions between two spatial points repress but do not eliminate the correlations.  (2) On the other hand, if the local competing interactions cancel and create a local minimum energy degeneracy, correlations occur only in the form of the local configurations that participate into the degenerate local minimum energy. This effect is called frustration \cite{Toulouse}.

Hierarchical models \cite{BerkerOstlund,Kaufman1,Kaufman2} are exactly solvable microscopic models that are widely used.\cite{Myshlyavtsev,Derevyagin,Shrock,Monthus,Sariyer,Ruiz,Rocha-Neto,Ma,Boettcher5}  The construction of a hierarchical model is illustrated in Fig.1(a) \cite{BerkerOstlund}.  The hierarchical lattice constructed in Ref.\cite{McKayChaos}, for the study of competing interactions, included both the repression and frustration effects explained above.  This construction is shown in Fig.1(b,c) and includes, in its basic graph, the specified indices of the number $p_b$ of frustrated units, the number $p$ of frustrating cross bonds, the number $p_c$ of repressed units, the lengths $m_1, m_2>m_1$ of the repressed chains.

Each line segment in Fig.1 represents a nearest-neighbor spin-spin interaction $Js_is_j$, where at each site $i$ of the lattice the Ising spin $s_i =\pm1$.  The Hamiltonian of the entire hierarchical lattice is
\begin{equation}
-\beta \mathcal{H}= J \sum_{\langle ij \rangle} s_is_j \,,
\end{equation}
where $\beta=1/kT$ and $\langle ij \rangle$ denotes summation over all nearest-neighbor site pairs. The exact renormalization-group solution of the hierarchical model proceeds in the direction reverse to its construction (Fig.1), by summing over the internal spins shown with the dark circles.  This summation yields the renormalization-group recursion relation $J'=J'(J)$, which becomes chaotic in wide ranges in the space of the indices $p,p_b,p_c,m_1,m_2=m_1+\Delta m$ chosen in the construction of the hierarchical model.  Fig.2 shows the chaotic regions in cross sections of this multidimensional space of indices. Note islands that exclude chaos, such as the crescent and star region in the upper-right panel and all the lower panels in Fig.2. As explained in Ref.\cite{McKayChaos}, chaotic renormalization-group trajectories signal the spin-glass phase, strong and weak correlations occurring in a random sequence at consecutive length scales and with extreme sensitivity to infinitesimal changes in external conditions such as temperature.\cite{Aral}  Chaotic renormalization-group trajectories and thus the spin-glass phase have subsequently been obtained in the approximate solution of systems with competing interactions, ferromagnetic versus antiferromagnetic interactions \cite{Ilker1,Ilker2,Ilker3} or left-chiral versus right-chiral interactions \cite{Caglar1,Caglar2,Caglar3}, in cubic-lattice systems and re-wired \cite{Ilker2} square-lattice systems.

\section{Chaos under Scale Change in Spin Glasses: A Controllable and Wide-Ranged Behavior for Projection from Complex Systems}

Multifractal spectra are calculated for chaotic renormalization-group trajectories of the standart chaotic hierarchical lattice (Fig.1), music tracks from different geographies and cultural trends (Table I), brain EEG signals under different conditions, and, for future work, any complex system data set, as explained in Appendix A. For a given set of spin-glass indices $p,p_b,p_c,m_1,m_2$, the starting temperature $T=J^{-1}$ of the chaotic renormalization-group trajectory is immaterial, as long as it is within the spin-glass phase, since a single chaotic trajectory is the asymptotic renormalization-group sink of the entire phase, attracting all initial conditions within the phase.  For example, numerically identical Lyapunov exponents $\lambda$ \cite{Collet,Hilborn}, quantifying the strength of chaos, are obtained for all initial conditions inside the phase:
\begin{equation}
\lambda = \lim _{n\rightarrow\infty} \frac{1}{n} \sum_{k=0}^{n-1}
\ln \Big|\frac {dJ_{k+1}}{dJ_k}\Big|\,,
\end{equation}
where $J_k$ is the value of the interaction constant in Eq.(1) after the $k$th renormalization-group transformation.  Thus, our chaotic trajectories are taken after throwing out the first 1000 iterations and then using the next $2^{13}$ iterations, to eliminate transient effects of the crossover to the sink.  In the examples shown in Fig.3, \textit{e.g.,} we see that there is much stronger chaos in the music ($\lambda = 0.305$) than in the brain signals ($\lambda = 0.040$).

As shown in Fig.3, the spin-glass multifractal spectrum points are fitted by a continuous quartic function.  The root-mean-square separation between the music (or brain EEG) multifractal spectrum points and the spin-glass continuous function is minimized by varying the original spin-glass indices of frustration number $p_b$ and strand length $m_1$.  At the end of this procedure, the statistical correlation coefficient $R_s$ between the spin-glass multifractal spectrum points and the spin-glass continuous function and the correlation coefficient $R_m$ between the music multifractal spectrum points and spin-glass continuous function $R_m$ together give the goodness of the fit between the spin-glass multifractal spectrum points and the music (or brain EEG) multifractal spectrum points. At the optimum fit, the calculated Lyapunov exponent $\lambda$ measures the strength of the fitting chaos.\cite{Gurleyen}

\section{Chaotic Spin-Glass Classification of Multicultural and Multigeographic Music}

Scatter plots of the fitted frustrated hierarchical Ising models to multicultural music tracks are displayed in Fig. 4. Each point represents the fitted model indices for a single music track. There are ten tracks for each of the eight genres, separately listed in Table I.  The genres are from Turkish music, namely Arabesque, Rap, Pop, Classical; and Western music, namely Blues, Jazz, Pop, Classical.  In each panel, for the indicated values of $p,p_c,\Delta m$, the values of the frustration number $p_b$ and repressed strand length $m_1$ are fitted to the music tracks.  Thus, each panel is an alternate attempt to resolve the same data of 80 music tracks.

Systematic patterns emerge and are quite surprising.  Firstly, essentially in all panels, the indices are organized in streaks that track each other. Blues, Jazz, and Western Classical (all from Western music), but not including Western Pop, stand apart.  The other grouping is Turkish music (Arabesque, Rap, Pop, Classical), but also including Western Pop, all well mixed.

\section{Chaotic Spin-Glass Classification of Brain EEG Signals from Differently Resting States and Cranular Regions}

Our calculated spin-glass multifractal indices for brain EEG signals, collected at different cranular probe locations and in the different states of music listening, resting with eyes open or closed, are displayed in Fig.5.\cite{Brain,BrainM,BrainMap}  Again, an even clearer systematic pattern emerges.  In the front cranular region, shown in the seven F panels, the indices fall on very well defined parallel streaks.  There is predominant overlap between EEG signals from music listening and resting with eyes closed, as opposed to resting with eyes open.  In the back cranular region, shown in the eight C panels, the separation is even clearer.  For each panel, \textit{i.e.,} each back cranular location, the spin-glass indices of the brain EEG data separate into two near-orthogonal branches.  The horizontal branch, namely the constant (low) frustration number $p_b$ branch, has the signals from music listening.  The near-vertical branch, namely the near-constant (low) strand length $m_1$, we find the EEG signals from music listening in its low $p_b$ segment, then the EEG signals from resting with eyes closed in the upper segment, and the EEG signals from resting with eyes open clustered at the top $p_b$ edge of the branch.

\section{Conclusion}
We have demonstrated that a complexity classification scheme can be developed from the fractal spectra of spin-glass chaos.  We examplified the procedure with 80 pieces multigeographic multicultural music from 8 genres and brain electroencephalogram signals.  Systematic patterns and, in music, an interesting group defection is detected.

The tailor-made breadth and exact solution of spin-glass chaos makes this procedure a very widely usable classification and analysis scheme for all sorts of complex data.

\begin{acknowledgments}
We thank Ratip Emin Berker and Deniz Ero\u{g}lu for very useful discussions and comments. Support by the Kadir Has University Doctoral Studies Scholarship Fund and by the Academy of Sciences of Turkey (T\"UBA) is gratefully acknowledged.
\end{acknowledgments}

\appendix

\section{Optimized Multifractal Sectra Fits between Chaotic Spin Glasses, Music, and Brain EEG}

Two types of data have been used, the first of which is the experimental data (music or brain EEG) and the other is the chaotic spin-glass data, from the exact renormalization-group solution of the frustrated Ising model on a hierarchical lattice. We first fix the $p, p_c, \Delta m = m_2 – m_1$ values and do the renormalization-group transformation for each $p_b$ and $m_1$ value between 0-200. While applying the renormalization-group transformations, we take $2^{13}$ iterations values after discarding the first 1000 iterations as crossover to asymptotic chaos. The occurrence of chaos for the given $p_b$ and $m_1$ values is checked by calculating the Lyapunov exponent. For chaos, we apply Chhabra-Jensen Algorithm \cite{ChhabraJensen} to obtain the multifractal spectrum $f(\alpha)$ from the recurring $J$ values. Thus, a repository of chaotic $p_b$ and $m_1$ values and their correspondent $f(\alpha)$ is created to fit the music and brain EEG data.

To prepare the music data for Chhabra-Jensen algorithm, firstly the audio files have been transformed into time series data. We choose the left-ear channel in the time series. Then we discard the first 50000 data, which correspond mostly to silence, and take the next $2^{\lfloor \log _2(N) \rfloor}$ data, where $N$ is the number of time series data points, because a power of 2 is needed in the Chhabra-Jensen algorithm.  This data is normalized as $x_i’=(x_i-\tilde{x}_i)/ \sigma_x$, where $\sigma_x$ is the root-mean-square deviation of the data points and $\tilde{x}_i = 1/(1+e^{x_i})$.

For the brain data, firstly the EEG data is imported to MATLAB(R2021a, The MathWorks, Inc., Natick) using FieldTrip Toolbox (https://doi.org/10.1155/2011/156869).\cite{EEG} Then, a bandpass filter is applied, a common procedure when dealing with EEG data. To eliminate possible artifacts caused by the filtering, the first and last 3000 data points are discarded. Again, the first $2^{\lfloor \log _2(N) \rfloor}$ of the data is used and is normalized as described above.

\section{Calculation of Multifractal Spectra of Chaotic Renormalization Group, Music, and Brain Signals}

The Chhabra-Jensen function written by Fran\c{c}a \textit{et al.} \cite{ChhabraJensen} is used, for all time-series data $A=\{a_{1},a_{2},...,a_{N}\}$ composed of $N$ data points. Firstly, we split the data into disjoint sub-blocks $B_i=\{b_{i1},b_{i2},...,b_{il}\}$ of length $l$. Then we calculate the consecutive probabilities of each block:
\begin{equation}
    P_i(l) = \frac{\sum_{j=1}^{l}b_{ij}}{\sum_{j=1}^{N}a_{j}}.
    \label{eqn:Probabilty}
\end{equation}
We normalize the $q$th power of probabilities $P_i$ and call it $\mu_i$:
\begin{equation}
    \mu_i(q,l) = \frac{{P_i(l)}^q}{\sum_{j}^{}{P_j(l)}^q}.
    \label{eqn:mu}
\end{equation}
Then we calculate the Hölder exponent $\alpha$ and the corresponding Hausdorff dimension $f(\alpha)$ to obtain the multifractal spectrum, by applying log-log fits to $M_\alpha(l)$ and $M_f(l)$:
\begin{equation}
\begin{aligned}
    M_\alpha(l)&= \sum_{i=1}\mu_i(q,l)log_{10}(P_i(l)),\\
    M_f(l)&=\sum_{i=1}\mu_i(q,l)log_{10}(\mu_i(q,l)).
\end{aligned}
   \label{eqn:Measures}
\end{equation}
The slopes found after regressing the log-log plots of $M_\alpha(l)$ and $M_f(l)$ are $\alpha$ and $f(\alpha)$.  Repeating this fit for different $q$ values, we obtain the multifractal spectrum.

\begin{table*}
\begin{tabular}{c c c c c
}
\hline
\vline &\textbf{Composition, Artist} &\vline  &\textbf{Composition, Artist} &\vline  \\
\hline
\hline

\vline &\textbf{Arabesque}   &\vline & \textbf{Blues}  &\vline  \\
\hline
\vline & Elimde Duran Foto\u{g}raf{\i}n, Bergen &\vline  & Lonely Bed, Albert Cummings  &\vline   \\
\hline
\vline & Sev Yeter, Bergen  &\vline & Lucille, B.B. King   &\vline  \\
\hline
\vline & K\"ullenen A\c{s}k, Cengiz Kurto\u{g}lu &\vline  & She Is Crazy, Coldfire  &\vline   \\
\hline
\vline & Can{\i}na Okuyaca\u{g}{\i}m, Ferdi Tayfur  &\vline & The Sky Is Crying, Coleman &\vline   \\
\hline
\vline & Vur Gitsin Beni, Ibrahim Tatl{\i}ses &\vline  & I Will Play the Blues for You, Daniel Castro  &\vline   \\
\hline
\vline & Bir Ate\c{s}e Att{\i}n, Kamuran Akkor &\vline  & I'd Rather Go Blind, Etta James  &\vline   \\
\hline
\vline & Kadehi \c{S}i\c{s}eyi K{\i}rar{\i}m Bug\"un, Kibariye &\vline  & The Bluesbreakers John Mayall &\vline    \\
\hline
\vline & Affet, M\"usl\"um G\"urses  &\vline & Last Two Dollars, Johnnie Taylor  &\vline   \\
\hline
\vline & Unutamad{\i}m, M\"usl\"um G\"urses &\vline  & Mannish Boy, Muddy Waters  &\vline   \\
\hline
\vline & Hatas{\i}z Kul Olmaz, Orhan Gencebay &\vline  & Feeling Good, Nina Simone  &\vline   \\
\hline
\hline
\vline &\textbf{Turkish Rap}   &\vline & \textbf{Jazz}  &\vline  \\
\hline
\vline & İzmir'in Ateşi, Ben Fero  &\vline & Collard Greens Gand Black Eyed Peas, Bud Powell  &\vline   \\
\hline
\vline & Ben Ağlamazken, Ceza &\vline  & More Today Than Yesterday, Charles Earland &\vline   \\
\hline
\vline & Med Cezir, Ceza &\vline & Take Five, Dave Brubeck  &\vline   \\
\hline
\vline & Suspus, Ceza &\vline & Take The A Train, Duke Ellington  &\vline  \\
\hline
\vline & Zenti, Ex &\vline & Song For My Father, Horace Silver  &\vline   \\
\hline
\vline & Bul Beni, Ezhel &\vline & My Favorite Things, John Coltrane &\vline   \\
\hline
\vline & Şehrimin Tadı, Ezhel  &\vline & Una Mas, Kenny Dorham  &\vline  \\
\hline
\vline & Baytar, Sagopa Kajmer  &\vline & The Sidewinder, Lee Morgan  &\vline  \\
\hline
\vline & Galiba, Sagopa Kajmer &\vline & Freddie Freeloader, Miles Davis  &\vline   \\
\hline
\vline & Makina, Uzi  &\vline & So What, Miles Davis  &\vline   \\
\hline
\hline
\vline &\textbf{Turkish Pop}   &\vline & \textbf{Western Pop}  &\vline  \\
\hline
\vline & Yerlerdeyim, Çelik &\vline  & Hello, Adele  &\vline  \\
\hline
\vline & Bi Daha Bi Daha, Demet Akalın &\vline & Faded, Alan Walker  &\vline  \\
\hline
\vline & Bangır Bangır, Gülşen &\vline & Wake Me Up, Avicii  &\vline   \\
\hline
\vline & Sahte, Hande Yener &\vline & Favorito, Camilo  &\vline  \\
\hline
\vline & Sevgilim, Murat Boz &\vline & Woman, Doja Cat  &\vline   \\
\hline
\vline & Yoksa Yasak, Oğuzhan Koç &\vline & Photograph, Ed Sheeran  &\vline  \\
\hline
\vline & Poşet, Serdar Ortaç  &\vline & Starving, Hailee Steinfeld  &\vline  \\
\hline
\vline & Şımarık, Tarkan &\vline & What Makes You Beautiful, One Direction &\vline   \\
\hline
\vline & Ebruli, Yaşar &\vline & Let Her Go, Passenger  &\vline  \\
\hline
\vline & Kalbim Tatilde, Ziynet Sali &\vline  &Closer, The Chainsmokers  &\vline   \\
\hline
\hline
\vline &\textbf{Turkish Classical}   &\vline & \textbf{Western Classical}  &\vline  \\
\hline
\vline & Hicaz Ud Taksimi &\vline & Toccata Och Fuga, Bach &\vline   \\
\hline
\vline & Karciyar Kanun Taksimi &\vline & Piano Concerto Nr.5 Allegro, Beethoven  &\vline  \\
\hline
\vline & Kürdî Ud Taksimi &\vline & Minuet, Boccherini  &\vline   \\
\hline
\vline & Kürdilî Hicazkar Ud Taksimi &\vline & Hungarian Dances, Brahms &\vline   \\
\hline
\vline & Muhayyer Kanun Taksimi &\vline & In The Hall of the Mountain King, Grieg  &\vline   \\
\hline
\vline & Nihavend Ud Taksimi &\vline & Ombra Mai Fu, Händel &\vline   \\
\hline
\vline & Rast Ud Taksimi &\vline  & Symphony Nr.40, Mozart &\vline   \\
\hline
\vline & Saba Kanun Taksimi &\vline & Factotum Aria from Barber of Sevilla, Rossini &\vline   \\
\hline
\vline & Suzinak Kanun Taksimi &\vline & Marche Militaire Nr.1, Schubert &\vline   \\
\hline
\vline & Uşşak Kanun Taksimi &\vline  & Piano Concerto No 1 B-Flat Minor, Tchaikovsky &\vline    \\
\hline
\hline

\end{tabular}
\caption{Spin-glass-chaos indexed (Fig.4) multicultural music compositions by genre, title, and artist.\\} \label{tab:1}
\end{table*}

\end{document}